\documentclass[12pt, twoside]{article}
\usepackage{a4wide,amssymb,cite,graphicx,subfigure}
\usepackage{epsfig}

\def\de{\partial}

\def\b{\beta}
\def\g{\gamma}

\def\d{\delta}

\def\s{\sigma}

\def\f{\phi}
\def\ep{\epsilon}

\newcommand{\Ref}[1]{(\ref{#1})}

\newcommand{\be}{\begin{equation}}
\newcommand{\ee}{\end{equation}}
\newcommand{\bea}{\begin{eqnarray}}
\newcommand{\eea}{\end{eqnarray}}
\newcommand{\beqar}{\begin{eqnarray*}}
\newcommand{\eeqar}{\end{eqnarray*}}

\newcommand{\ie}{{\it i.e.}\ }

\begin{document}

\baselineskip=16pt   
\begin{titlepage}  
\rightline{hep-th/0407208}  
\rightline{July 2004} 
\begin{center}  
  
\vspace{3cm}  
  
\large {\bf Brane solutions of a spherical sigma  model in six dimensions}  

\vspace*{1cm}  
\normalsize  
   
{\bf Hyun Min Lee\footnote{minlee@th.physik.uni-bonn.de} and  Antonios   Papazoglou\footnote{antpap@th.physik.uni-bonn.de}}

\medskip   
\medskip   
{\it Physikalisches Institut der Universit\"at Bonn}\\  
{\it Nussallee 12,  D-53115 Bonn, Germany}  

\vskip3cm \end{center}  
   
\centerline{\large\bf Abstract}

We explore solutions of six dimensional gravity coupled to a non-linear sigma model, in the presence of co-dimension two branes. We investigate the compactifications induced by a spherical scalar manifold and analyze the conditions under which they are of finite volume and singularity free. We discuss the issue of single-valuedness of the scalar fields and provide some special embedding of the scalar manifold to the internal space which solves this problem. These brane solutions furnish some self-tuning features, however they do not provide a satisfactory explanation of the vanishing of the effective four dimensional cosmological constant. We discuss the properties of this model in relation with the  self-tuning example based on a hyperbolic sigma model.

\vspace*{1mm}   

\end{titlepage}

\section{Introduction}

Recently, there has been a lot of work on extra dimensional models with brane sources in relation with the cosmological constant problem \cite{selftuning}. The aim has been to find solutions with zero effective four dimensional cosmological constant regardless of the value of the brane vacuum energy. This adjustment mechanism has been called self-tuning and is particularly promising for codimension-two branes \cite{Luty}. These branes have the property that they do not curve the extra space but only induce a conical deficit in the internal geometry. Thus, it is conceivable that the brane vacuum energy in this case is absorbed in a change of the deficit angle without affecting the properties of the bulk solution.

The latter scenario was studied in detail in the framework of compactifications in the presence of gauge field fluxes. This kind of compactification was first considered in the seventies \cite{Scherk} (under the name of spontaneous compactifications) and have been revisited recently because of the property to fix some or all of the moduli of extra dimensional models. The first attempt to realize self-tuning in flux compactifications was done by \cite{Carroll} where an example of a ``rugby-ball''-shaped internal space was constructed and was shown that flat solutions existed for any value of the brane tension. The model included a bulk fine tuning between the flux and the bulk cosmological constant which can be relaxed if supersymmetry is invoked in the bulk \cite{Quevedo} (generalizing the supergravity solution of \cite{Salam}). However, it was soon realized that due to flux quantization \cite{Navarro} (or even flux conservation \cite{fluxconserv}), a relation between the brane tension and the bulk cosmological constant is introduced and so the self-tuning is ruined. For a detailed discussion of properties of these models (with or without supersymmetry) see \cite{Cline,Gibbons,Tasinato,Peter,HM,Graesser}.

There is another mechanism that induces compactifications of the extra space dimensions which utilizes a non-linear sigma model and has not been discussed in such a detail as the previous case. This kind of compactification was first considered in the eighties \cite{Zwiebach1,Zwiebach2} in the framework of supergravity and used non-trivial backgrounds of the fields of a  hyperbolic sigma model to compactify the internal space to a manifold called ``tear-drop''. Recently, this kind of solution was generalized by \cite{Kehagias} with codimension-two branes, yielding self-tuning solutions which do not have the complications of the previous flux models. These models, however, possesses a naked singularity in the bulk and although one can prevent energy flow in the singularity by appropriate boundary conditions \cite{Zwiebach2}, the solution cannot be trusted close to the singularity since the curvature becomes significant.

In the present paper, we consider instead a spherical sigma model and derive solutions with  codimension-two  branes. We first present the analog of the ``rugby-ball'' compactification and then generalize to more general compactifications with azimuthal symmetry. Depending on values of the sigma model coupling and the brane tensions, we can find non-singular solutions of finite volume. We discuss the issue of the single-valuedness of the sigma model fields and some special embedding of the scalar manifold to the internal space which circumvents this problem. We note that although the above solutions have self-tuning features, they cannot provide a satisfactory explanation to the cosmological constant problem since there exist nearby curved solutions for non-zero bulk cosmological constant. Finally, we compare this model with the self-tuning model based on a hyperbolic sigma model and conclude.

\section{Model setup}

We will consider a six-dimensional model with gravity, a  two-dimensional non-linear sigma model with metric $f_{ij}(\f)$ in the presence of codimension-2 branes. The full action of the system is
\be
S=\int d^6x \sqrt{-g}\bigg(\frac{1}{2}R
-\frac{1}{2}k f_{ij}\partial_M\phi^i\partial^M\phi^j\bigg)
+S_4, \label{action}
\ee
where $\phi^i(i=1,2)$ are real scalar fields, $k>0$\footnote{We restrict the sign of $k$ in order not to have a ghost-like kinetic term for the sigma model.}  is the coupling of the sigma model to gravity. The scalar manifold is chosen to be a sphere with  metric $f_{ij}$ given by
\be
d\s_f^2= (d\f^1)^2 + \sin^2\phi^1 (d\f^2)^2 . \label{sigmetric}
\ee
The brane action $S_4$ is given by the following localized terms
\be
S_4=\sum_{i=1,2} \int d^4xd^2y\sqrt{-g^{(i)}}(-\Lambda_i)
\delta^2(y-y_i) ,
\ee
where $g^{(i)}_{\mu\nu}$, $\Lambda_i$ and $y_i$
are brane-induced metrics, brane tensions and positions of the branes 
in extra dimensions, respectively.

We wish to find solutions of the above system where the internal two dimensional manifold is compactified  and is axisymmetric. The two 3-branes are placed at antipodal points on the axis of symmetry of the internal manifold.

The metric variation of the  above action (\ref{action}) 
gives rise to the Einstein equation 
\begin{eqnarray}
R_{MN}-\frac{1}{2}g_{MN}R
=k~f_{ij}\bigg(\partial_M\phi^i\partial_N\phi^j
-{1 \over 2}g_{MN}\partial_P\phi^i\partial^P\phi^j\bigg)
+T^{(4)}_{MN} ,
\end{eqnarray}
with the brane energy momentum tensor
\begin{eqnarray}
T^{(4)}_{MN}=-\sum_{i=1,2}\frac{\sqrt{-g^{(i)}}}{\sqrt{-g}}
\Lambda_i g^{(i)}_{\mu\nu}\delta^\mu_M\delta^\nu_N\delta^2(y-y_i).
\end{eqnarray}
We can rewrite the Einstein equation in a simpler way in terms of the Ricci tensor as
\begin{eqnarray}
R_{MN}=k~f_{ij}\partial_M\phi^i\partial_N\phi^j
-\sum_{i=1,2}\frac{\sqrt{-g^{(i)}}}{\sqrt{-g}}
\Lambda_i(\delta^\mu_M\delta^\nu_N g^{(i)}_{\mu\nu}-g_{MN})\delta^2(y-y_i). 
\label{ricci}
\end{eqnarray}

On the other hand, the field equation for the scalars is
\begin{eqnarray}
\frac{2}{\sqrt{-g}}\partial_M\bigg(\sqrt{-g}~
f_{ij}\partial^M\phi^j\bigg)
=\frac{\partial f_{kl}}{\partial\phi^i}
\partial_M\phi^k\partial^M\phi^l  ,
\end{eqnarray}
or equivalently
\be
\Box \f^i=\frac{1}{\sqrt{-g}}\partial_M\bigg(\sqrt{-g}~\partial^M\phi^i\bigg)
=-\g^i_{kl}~\de_M \f^k \de^M \f^l , \label{scalar}
\ee
where $\g^i_{kl}$ are the Christoffel symbols for the sigma model metric  $f$.

\section{``Rugby-ball''-shaped internal space} 

In order to find a background solution in this model,
let us take the ansatz for the metric with factorizable extra dimensions as
\begin{eqnarray}
ds^2=g_{\mu\nu}(x)dx^\mu dx^\nu+\gamma_{mn}(y)dy^mdy^n ,
\end{eqnarray}
where $g_{\mu\nu}(x)$ denotes the four dimensional spacetime which is taken to be maximally symmetric. Then, when the scalars depend
only on extra coordinates, eq.~(\ref{ricci}) implies
that $R_{\mu\nu}=0$, \ie the only maximally symmetric spacetime that is a solution is Minkowski spacetime.

With the 4d flat metric, $g_{\mu\nu}=\eta_{\mu\nu}$, let us take the ans\"atze
as following
\begin{eqnarray}
ds_{\g}^2&=&R^2_0(d\theta^2
+\beta^2\sin^2\theta d\psi^2),\\ 
\phi^1&=&\theta, \\ 
\phi^2&=&\beta\psi+c , \label{csph}
\end{eqnarray}
with $c$ being an integration constant.
Then, the field equation for the scalars (\ref{scalar}) are satisfied
while the Einstein equation (\ref{ricci}) in the bulk is also satisfied
only for $k=1$. We note that the radius of extra dimensions $R_0$ is not determined 
from the equations of motion. 
Matching the singular terms coming from the conical
singularities with the brane source terms leads to the following relation between  the brane tensions and the parameter $\b$
\begin{eqnarray}
\Lambda_1=\Lambda_2=2\pi(1-\beta),
\end{eqnarray}
with the deficit angle of the two branes being equal to $\d=2\pi(1-\b)$. Therefore, we find that changing the brane tensions is compensated 
by the deficit angle on the sphere.
To avoid the tuning between brane tensions, we need to consider the orbifold
$S^2/Z_2$ with $Z_2$ acting on the sphere coordinates as
\be
\theta\rightarrow \pi-\theta, \ \ \ \psi\rightarrow \psi.
\ee
Then, in order to have a well defined solution,  it is enough to impose the $Z_2$ parities on $\phi^1$ and $\phi^2$ as
\begin{eqnarray}
\phi^1\rightarrow \pi-\phi^1, \ \ \ \phi^2\rightarrow\phi^2.
\end{eqnarray}

\section{General internal space} \label{general}

In the previous section we obtained a flat solution with ``rugby-ball''-shaped extra dimensions for which 
two brane tensions are the same. This happens  only for $k=1$.  This is a 
 sort of a bulk fine-tuning of the scalar coupling to gravity. 
In this section, we find more general solutions with different embedding of the brane singularities 
in the metric in which one doesn't need to have $k=1$.

Let us define a complex scalar field in terms of the  $\f^i$'s  as
\begin{eqnarray}
\Phi=\bigg(\tan\frac{\phi^1}{2}\bigg) e^{i\phi^2}.
\end{eqnarray}
Then, the scalar manifold metric in eq.\Ref{sigmetric} becomes
\be
d\s_f^2=\frac{4~d\Phi d{\bar\Phi}}{(1+|\Phi|^2)^2}.
\ee
Then, with the above definition of fields, the Einstein equations \Ref{ricci} and the field equation for $\Phi$ \Ref{scalar}, are
respectively as following
\begin{eqnarray}
R_{MN}&=&\frac{2k}{(1+|\Phi|^2)^2}(\partial_M\Phi\partial_N{\bar\Phi}
+\partial_N\Phi\partial_M{\bar\Phi}) \nonumber \\
&-&\sum_{i=1,2}\frac{\sqrt{-g^{(i)}}}{\sqrt{-g}}
\Lambda_i(\delta^\mu_M\delta^\nu_N g^{(i)}_{\mu\nu}-g_{MN})\delta^2(y-y_i),
\end{eqnarray}
and
\begin{equation}
\frac{1}{\sqrt{-g}}\partial_M(\sqrt{-g}\partial^M\Phi)
=\frac{2{\bar\Phi}}{1+|\Phi|^2}\partial_M\Phi\partial^M{\Phi}. \label{sceq1}
\end{equation}

In order to find a flat solution, let us assume that extra dimensions are 
factorized and take the ans\"atze for the internal metric and the complex scalar field 
in complex coordinates as 
\begin{eqnarray}
ds^2_2&=&r^2_0 e^{2A(z,{\bar z})}dzd{\bar z}, \\
\Phi&=&\Phi(z,{\bar z}).
\end{eqnarray}
where the ``radius'' $r_0$ is a scale typical of the size of the internal space.  Then, the ($z\bar{z}$) Einstein equation and the field equation are
\begin{eqnarray}
-2\partial{\bar\partial}A&=&\frac{2k}{(1+|\Phi|^2)^2}
(\partial\Phi{\bar\partial}{\bar\Phi}+{\bar\partial}\Phi\partial{\bar\Phi})
+\sum_{i=1,2}\Lambda_i  \delta^2(z-z_i), \\
\partial{\bar\partial}\Phi
&=&\frac{2{\bar\Phi}}{1+|\Phi|^2}\partial\Phi{\bar\partial}\Phi. \label{sceq2}
\end{eqnarray}

The ($zz$) Einstein equation dictates that $\Phi$ is either holomorphic or antiholomorphic. Then the scalar field equation is automatically satisfied for any (anti)holomorphic function $\Phi=\Phi(z)$ ($\Phi=\Phi(\bar{z})$). Assuming that one of the branes is located at $z_1=0$,  we can readily get the solution for the metric in terms of
the scalar field as
\begin{eqnarray}
A=-k\ln(1+|\Phi|^2)-a\ln|z|+f(z)+{\bar f}({\bar z}),
\end{eqnarray}
where the functions $\Phi(z)$, $f(z)$ and ${\bar f}({\bar z})$ are regular at $z=0$.

At this point, in order to illustrate some explicit solutions,  we take a simple holomorphic function for the complex scalar
\begin{eqnarray}
\Phi(z)=c_0 z^b, \label{cgen}
\end{eqnarray}
with $c_0$ a phase (\ie $|c_0|=1$) and $b$. Then the internal metric becomes
\begin{eqnarray}
ds^2_2=r^2_0 |z|^{-2a}\frac{dzd{\bar z}}{(1+|z|^{2b})^{2k}}.
\end{eqnarray}

In order to see how the parameter $a$ is related to the tension of the brane sitting at $z=0$, we examine the metric at the origin
\be
ds^2_2=r^2_0[d\rho^2+(1-a)^2\rho^2 d\psi^2],
\ee
where a change of coordinates $\rho=|z|^{1-a}/(1-a)$ has been performed. Then we find that the  conical singularity at $z=0$ must be matched with the brane tension as
\be
{\Lambda_1 \over 2\pi}=1-|1-a| \equiv 1-\b_1,
\ee
and the deficit angle of the brane sitting at $z=0$ is $\d_1=2\pi (1-\b_1)$. As we will see shortly the condition for finite volume of the internal space forces $a<1$, so finally
\begin{eqnarray}
\frac{\Lambda_1}{2\pi} =a \equiv  1-\b_1.\label{brane1}
\end{eqnarray}

As we see from the metric, the antipodal point of $z=0$ on the axis of symmetry of the internal space is $z \to \infty$ (note that this point is at finite proper distance from $z=0$). At this point we should put in principle a second 3-brane.  In order to see how the parameter $b$ is related to the tension of the brane sitting at $z \to \infty$, we examine the asymptotic form of the metric
\be
ds^2_2=r^2_0[d\rho^2+(1-a-2kb)^2\rho^2 d\psi^2],
\ee
where a change of coordinates $\rho=|z|^{1-a-2kb}/(1-a-2kb)$ has been performed. Then, we find that the additional conical singularity at $z \to \infty$
must be matched with the other brane tension as
\be
{\Lambda_2 \over 2\pi}=1-|1-a-2kb| \equiv 1-\b_2,
\ee
and the deficit angle of the brane sitting at $z \to \infty$ is $\d_2=2\pi (1-\b_2)$. As we will see shortly the condition for finite volume of the internal space forces $a+2kb>1$, so finally
\be
{\Lambda_2 \over 2\pi}=2-a-2kb \equiv 1-\b_2. \label{brane2}
\ee

Using eq.~(\ref{brane1}), we can rewrite the above condition as
\begin{eqnarray}
\Lambda_1+\Lambda_2=4\pi (1-kb).\label{brane2a}
\end{eqnarray}
Therefore, in view of eqs.~(\ref{brane1}) and (\ref{brane2a}), 
we find that any change of brane tensions can be compensated 
via $a$ and $b$ which are parameters in the solutions, 
maintaining the flat solution. In the following we will assume that $b>0$ since the $b<0$ case is its dual by changing $z \to 1/z$.

The size of extra dimensions $r_0$ is not determined, just as in the previous section. Therefore, there exists a modulus in the system which corresponds to the volume.

The solution we have obtained upto now could suffer in principle from curvature singularities at $z=0$ or $z \to \infty$. We should make sure that first of all the following curvature invariants (computed for the specific background) are everywhere finite
\be
R^2=2R_{MN}^2=R_{MNK \Lambda}^2={64b^4k^2 \over r_0^4}r^{4(a+b-1)}(1+r^{2b})^{4(k-1)} ,
\ee
where we have set $z=r~e^{i\psi}$. Thus, to have regular solutions the following conditions should be satisfied
\be
a+b   \geq 1~~~~~,~~~~~a+b(2k-1) \leq 1 . \label{condcur}
\ee

Additionally, we wish that the volume of the internal space is finite. This is proportional to 
\be
\int_0^{\infty}dr{r^{1-2a} \over (1+r^{2b})^{2k}}.
\ee
Thus, to have finite volume solutions the following conditions should be satisfied
\be
a<1~~~~~,~~~~~a+2kb>1 . \label{condvol}
\ee

Let us discuss now some special points in the above allowed parameter space.

\begin{enumerate}

\item[{\bf i.}] For ${\bf k=1}$ we see that the only way to satisfy all the constraints is when $a+b=1$ and $a<1$. Then $\Lambda_1=\Lambda_2=2\pi a$ and we obtain exactly the solution of the previous section of the ``rugby-ball''-shaped internal space. The metric is then indeed written as
\be
ds^2_2=r^2_0 \frac{dzd{\bar z}}{(|z|^a+|z|^{2-a})^{2}}=r^2_0 \frac{dzd{\bar z}}{|z|^2(|z|^\b+|z|^{-\b})^{2}} ,
\ee
which is the metric of the ``rugby-ball'' as in Ref.\cite{Carroll} and $r_0=2R_0\b$ where $R_0$ the radius of the ``rugby-ball''. The case  $a=0$, $b=k=1$ corresponds to the sphere.

\item[{\bf ii.}] For ${\bf bk=1}$ we see from \Ref{brane2a} that $\Lambda_1+\Lambda_2=0$ so the two branes have opposite tensions and the geometry resembles a ``heart''-shaped internal space, as in Fig.\ref{heartg}. From the constraints \Ref{condcur}, \Ref{condvol}, we obtain
\be
a+b \geq 1~~~~~,~~~~~a \leq b-1~~~~~,~~~~~-1<a<1 .
\ee
The allowed parameter space is shown in Fig.\ref{heartp}. Let us note that for the semi-line $a=0$, $b>1$ we obtain a configuration without branes which is an ellipsoid. The point $a=0$, $b=k=1$ corresponds to the sphere.

\begin{figure}[t]
\begin{center}
\epsfig{file=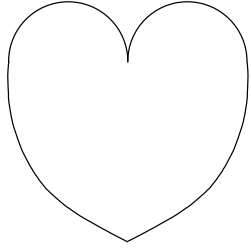,width=3.25cm}
\caption{The ``heart''-shaped geometry with the parameters of case ii.}
\label{heartg}
\end{center}
\end{figure}

\begin{figure}[h]
\begin{center}
\epsfig{file=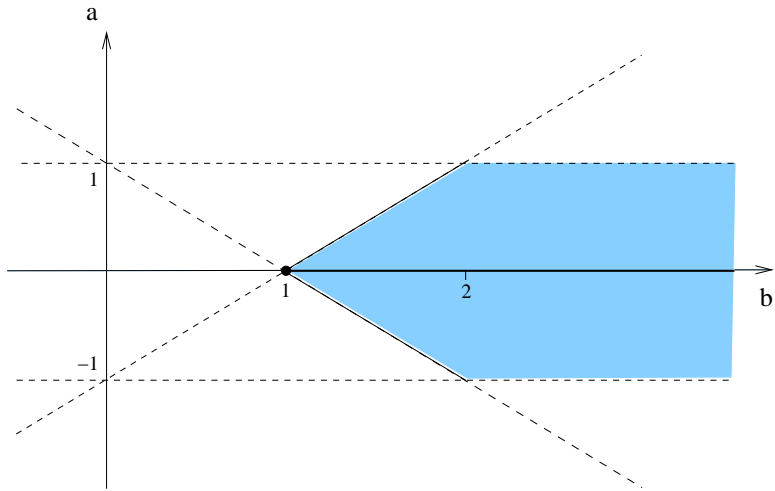,width=10.25cm}
\caption{
``Heart''-shaped solution: The allowed parameter space of $a,b$ for $kb=1$.
The dot point and the solid line in the shaded region
correspond to a sphere and an ellipsoid without branes, respectively.}
\label{heartp}
\end{center}
\end{figure}

\item[{\bf iii.}] For ${\bf a=0,~bk \neq 1}$ or for ${\bf a=2(1-kb),~ bk \neq 1}$ the internal space supports only one three-brane as in Fig.\ref{rheartg}. These two cases are related by duality $z \to 1/z$, so let us examine only the case with $a=0$, $bk \neq 1$ where $\Lambda_1=0$ and $\Lambda_2=4\pi (1-kb)$. Then the constraints \Ref{condcur}, \Ref{condvol} give
\begin{figure}[t]
\begin{center}
\epsfig{file=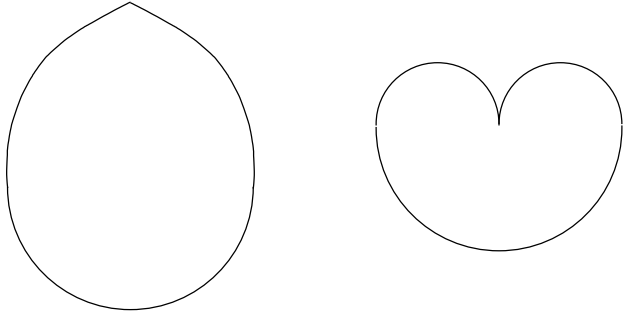,width=8.25cm}
\caption{The one three-brane geometries with the parameters of case iii.
The ``drop''-shaped (round ``heart''-shaped) corresponds to a positive (negative) brane tension.
}
\label{rheartg}
\end{center}
\end{figure}
\be
b \geq 1~~~~~,~~~~~bk>1/2~~~~~,~~~~~b(2k-1)\leq 1 .
\ee
The allowed parameter space is shown in Fig.\ref{rheartp}. Let us note that $\Lambda_2$ can have both signs depending on $b$.

\begin{figure}[h]
\begin{center}
\epsfig{file=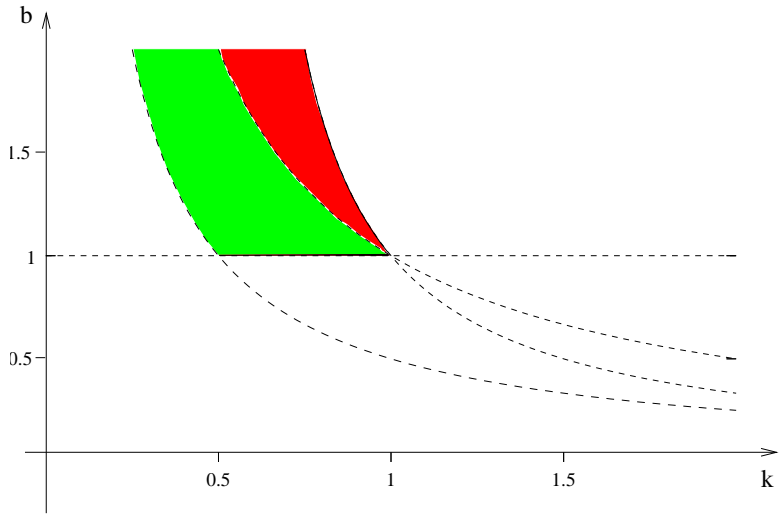,width=10.25cm}
\caption{
One three-brane solution: The allowed parameter space of $b,k$
for $a=0$ and $kb\neq 1$. The brane tension is positive for the green region
while it is negative for the red region.
}
\label{rheartp}
\end{center}
\end{figure}

\end{enumerate}

Apart from the above special cases, it is easy to see that there exist generic regions in the $(a,b,k)$  parameter space where the conditions \Ref{condcur}, \Ref{condvol} are satisfied. The important observation is that  these allowed regions are not isolated points in the parameter space but rather continuous intervals. So the parameters are allowed to vary continuously  without affecting the flatness of the solution.

\begin{figure}[t]
\begin{center}
\epsfig{file=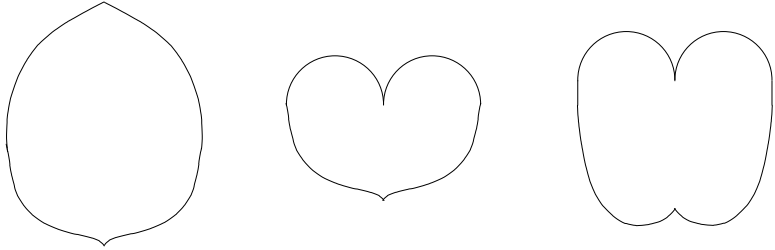,width=10.25cm}
\caption{The possible  geometries with $k=1/2$.
The figures depict two positive tensions, opposite tensions and 
two negative tensions, in order from the left to the right.
}
\label{geng}
\end{center}
\end{figure}

\begin{figure}[ht]
\begin{center}
\epsfig{file=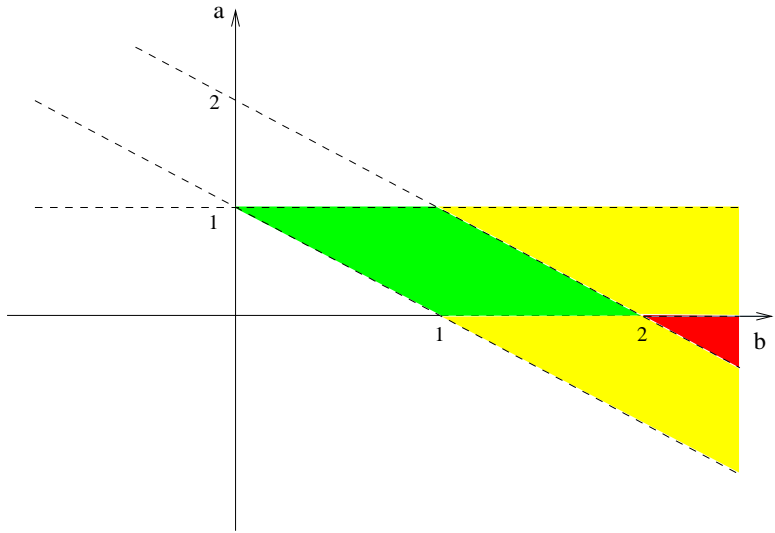,width=10.25cm}
\caption{
The allowed parameter space of $a,b$ for $k=\frac{1}{2}$. Both
brane tensions are positive (negative) for the green (red) region
while brane tensions take opposite signs for the yellow region.
}
\label{genp}
\end{center}
\end{figure}

As an illustration of the above remark let us consider a specific example with  $k=1/2$. Then from \Ref{condcur}, \Ref{condvol}   we have
\be
a<1~~~~~,~~~~~a+b>1 .
\ee
The possible geometries are shown in Fig.\ref{geng} and the allowed parameter space in Fig.\ref{genp}.

\section{Single-valuedness of the scalar field} \label{single}

In finding the above solutions we have omitted checking whether the scalar field is single-valued. In that sense all the above solutions are incomplete. In this section we will show that generically there exists a problem which can be solved by appropriate embedding of the scalar manifold coordinates to the internal space geometry. 

Let us first note the problem: The scalar field should be single valued once we perform a $2\pi$ rotation (of $\psi$) around the axis of symmetry\footnote{We thank Stefan F\"orste for discussions on this point.}. In other words
\be
\Phi(r,0)=\Phi(r,2\pi) ~~~~~ \Rightarrow ~~~~~ e^{2\pi b i}=1~~~~~ \Rightarrow ~~~~~ b=n,~~~~n={\rm integer}.
\ee
But the parameter $b$ is not generically an integer (since it is related to the tension of one of the branes), so one is directed to identifying points of the scalar manifold in order that the $\Phi$ field to be periodic. The latter  amounts, however, to changing the scalar manifold and thus the dynamics of the system every time the brane tensions change. This is a mere fine-tuning  which we want  to avoid if we wish to use the previous  solutions for obtaining self-tuning.

To circumvent this problem, we need to embed the scalar manifold in the internal space in a more contrived way. For this purpose we define the new field $X$ instead of $\Phi$
\be
X=\bigg(\tan\frac{\phi^1}{2}\bigg) e^{iK(\phi^2)},
\ee
where $K(\phi^2)$ is a function which is to be determined by the requirement that the scalar field $X$ is single-valued. The latter condition reads
\be
K[\phi^2(2\pi)]=K[\phi^2(0)]+2\pi n , \label{cond}
\ee
where $n$ is an integer.  Remember also that for our solutions $\phi^2=b \psi +c$ where $c$ is the integration constant appearing in \Ref{csph} and in \Ref{cgen} if we set $c_0=e^{ic}$. The trivial choice $K(\phi^2)=\phi^2$ fails to give an equation which determines $c$ and instead quantizes $b$. But for generic choices of $K(\phi^2)$, it is possible to have arbitrary (and continuous) $b$ and satisfy the single-valuedness condition by choosing an appropriate integration constant $c$.
 
Let us discuss a simple example of embedding which serves the above purpose
\be
K(\phi^2)=\phi^2+\ep (\phi^2)^2 ,
\ee
where $\ep$ is a parameter characteristic of the embedding. Then the condition \Ref{cond} gives
\be
c=-\pi b+{1 \over 2 \ep}\left({n \over b}-1\right).
\ee
For this choice of the integration constant $c$, the scalar field $X$ is rendered single valued. 


Finally, let us get an better insight in the special embedding that we have chosen. The redefinition of fields gives a mapping $X=f(\Phi,\bar{\Phi})$, so it gives a solution for $X$ which is non-holomorphic. What we have actually done by passing from the (multi-valued) field $\Phi$ to the (single-valued) $X$, is to find a solution of the equations of motion for non-holomorphic embeddings keeping the {\it same} solution for the spacetime metric.

\section{Nearby curved solutions}

Up to now, we assumed that there is no bulk cosmological constant. 
In this section we consider the model with non-zero bulk cosmological constant. We show that there exist nearby curved solutions for the flat solution
of the ``rugby-ball''-shaped internal space.

When we add a nonzero bulk cosmological 
constant $\Lambda_b$ to the sigma model action, the bulk action becomes 
\begin{eqnarray}
S_{bulk}=\int d^6 x \sqrt{-g}\bigg(\frac{1}{2}R-\Lambda_b
-2 k\,\frac{\partial_M\Phi\partial^M{\bar\Phi}}{(1+|\Phi|^2)^2}\bigg).
\end{eqnarray}
Then, the modified Einstein equation is 
\begin{eqnarray}
R_{MN}&=&\frac{1}{2}\Lambda_b\, g_{MN}
+\frac{2k}{(1+|\Phi|^2)^2}(\partial_M\Phi\partial_N{\bar\Phi}
+\partial_N\Phi\partial_M{\bar\Phi}) \nonumber \\
&-&\sum_{i=1}^2\frac{\sqrt{-g^{(i)}}}{\sqrt{-g}}
\Lambda_i(\delta^\mu_M\delta^\nu_N g^{(i)}_{\mu\nu}-g_{MN})\delta^2(y-y_i),
\end{eqnarray}
and the field equation for $\Phi$ is the same as eq.~\Ref{sceq1}.

\begin{figure}[t]
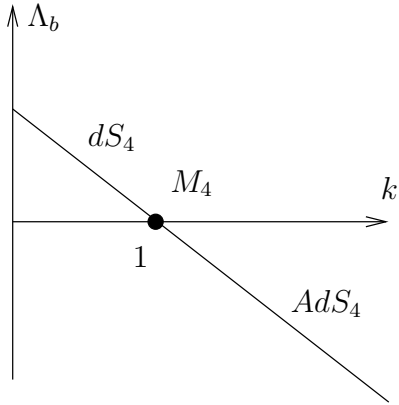

\center\input cosm.pstex_t
\caption{The bulk cosmological constant $\Lambda_b$ as a function of $k$. The four dimensional Minkowski space (dot) is continuously connected to dS and AdS space.}
\label{cosmol}
\end{figure}

Let us take the metric ansatz with factorized extra dimensions as 
\begin{eqnarray}
ds^2=g_{\mu\nu}(x)dx^\mu dx^\nu+r^2_0 e^{2A(z,{\bar z})}dz d{\bar z} ,
\end{eqnarray}
where $g_{\mu\nu}(x)$ denotes the four dimensional maximally symmetric spacetime
with its Ricci tensor given by 
$R_{\mu\nu}=3\lambda g_{\mu\nu}$. 
Here, $\lambda$ is a constant parameter 
which gives a 4d dS solution for $\lambda>0$,
a 4d flat solution for $\lambda=0$, and a 4d AdS solution for $\lambda<0$.  
Then, the Einstein equations give rise to
\begin{equation}
\lambda=\frac{1}{6}\Lambda_b,
\end{equation} 
and 
\begin{equation}
-2\partial{\bar\partial}A=\frac{1}{4}\Lambda_b\, r^2_0\, e^{2A}
+\frac{2k}{(1+|\Phi|^2)^2}
(\partial\Phi{\bar\partial}{\bar\Phi}+{\bar\partial}\Phi\partial{\bar\Phi})
+\sum_{i=1,2}\Lambda_i \delta^2(z-z_i).\label{extracc}
\end{equation} 
The field equation for the scalar field is the same as eq.~\Ref{sceq2}. As in the flat case, the ($zz$) Einstein equation dictates that $\Phi$ is (anti)holomorphic, and for any such function the scalar field equation is trivially satisfied. Then, taking the solution for the metric as
\begin{eqnarray}
A=-\ln(1+|\Phi|^2)-a\ln|z|, \label{assume}
\end{eqnarray}
we find from eq.~(\ref{extracc}) the holomorphic solution 
for $\Phi$ as follows,
\begin{eqnarray}
\Phi=c_0 z^{1-a} ,
\end{eqnarray}
with $c_0$ a constant phase. The radius of the sphere is 
 determined by relation with $k$ and $\Lambda_b$ as
\begin{eqnarray}
r^2_0=\frac{8(1-k)(1-a)^2}{\Lambda_b}.
\end{eqnarray} 
The parameter $a$ is related as usual to the brane tension as 
\begin{eqnarray}
a=\frac{\Lambda_1}{2\pi}=\frac{\Lambda_2}{2\pi}.
\end{eqnarray}
Therefore, the curved solutions for a nonzero $\Lambda_b$ and $k\neq 1$,
 are continuously connected to the ``rugby-ball''-shaped flat solution 
with $\Lambda_b=0$ and $k=1$ (see Fig.\ref{cosmol}). 
Note that for $0<k<1$, there exist only dS solutions while for $k>1$, 
there exist only AdS solutions.  

We should note here that the above is a simple solution by assuming the ansatz  \Ref{assume} and that there could  exist in general solutions with complicated embeddings of $\Phi$ if we change the ansatz \Ref{assume}. These more general solutions could provide nearby curved analogues for the other non-constant curvature compactifications that we found in section \ref{general}.

\section{Discussion and conclusions}

In this paper we have explored brane solutions for the spherical sigma model. It is helpful to recall what are the analogous brane solutions of the hyperbolic sigma model \cite{Kehagias} and compare them. The first important difference between the two approaches has to do with supersymmetry. In the hyperbolic case, the sigma model arises in six dimensional supergravity and thus the fine-tuning of the bulk cosmological constant can be explained. We should note here that the model as it stands, with vacuum branes, is supersymmetric and thus there is no mystery on why Weinberg's theorem is not applicable\footnote{We thank Hans-Peter Nilles and  Gianmassimo Tasinato for discussions on this point.}. [This also raises a question about the behaviour of the system regarding the effective cosmological constant if supersymmetry is broken.]  On the other hand, the spherical sigma model cannot arise in supergravity and thus one has no justification on setting the bulk cosmological constant to zero.
 
A second important difference between these models has to do with the presence of a naked singularity in the hyperbolic case\footnote{We note that for $k=-1$ (when the scalars have ghostlike kinetic term), one can see that the singularity is absent but the volume of the internal space diverges.}. There have been arguments in \cite{Zwiebach2,Kehagias} that with appropriate boundary conditions on the naked singularity one can prevent energy, angular momentum and $U(1)$ charge (related to the azimuthal isometry) to flow in the  singularity. However, the solution remains troublesome because infinitesimally close to the singularity the curvature explodes and the description of the theory breaks down. It is not clear if the completion of the theory in that high curvature regime will retain the properties discussed before. On the other hand, the spherical sigma model has no singularity problem for certain intervals of the brane tensions and sigma model coupling and thus is completely under control in the present theory. 

Finally, a third difference between the two models is the issue of the single-valuedness of the scalar fields. In the hyperbolic case there is again a problem of single-valuedness which however can be solved easily by having $b=n$, $n=$integer. Note that without this condition one is forced to identify points of the scalar manifold in such a way that the scalar gets single valued, which as discussed in section  \ref{single} amounts to fine-tuning. However, having $b$ to be an integer poses no problem as in the spherical case. In the spherical case the parameter $b$ was related to the tension of the second three-brane and thus could not be in general an integer. On the contrary, in the hyperbolic case there is no relation of $b$ with other input quantities and thus a solution with $b=$integer is satisfactory, without any need of a special embedding of the scalar manifold to the internal space.

Let us now briefly comment about the moduli of the system of the spherical sigma model (similar conclusions are expected to hold also for  the hyperbolic case). As was seen in the previous sections, the ``radius'' $r_0$ of the internal space is undetermined for the flat space compactifications. Thus there exists a massless scalar in the four dimensional theory which should be stabilized by some mechanism. There is no guarantee that the stabilization mechanism will not disturb the self-tuning property. On the other hand, for curved vacua like the ones of the previous section, the ``radius'' $r_0$ of the internal space is fixed and the radion is massive, with mass that depends on the effective four dimensional cosmological constant.

In conclusion,  we have presented new solutions of a six dimensional non-linear sigma model in the presence of codimension-two branes. We discussed in detail the conditions that should be satisfied for absence of singularities and for obtaining an internal space of finite volume. We noted that there is parameter space where there exist one or two  branes along the axis of symmetry and that they can have any combination of tension (\ie positive or negative). In order for the solutions to be single valued, one has to use a special embedding of the scalar manifold to the internal space. These solutions furnish some self-tuning feature, in the sense that the brane vacuum energy is not related to the flatness of the four dimensional effective theory. However, these solutions cannot still give a satisfactory resolution of the cosmological constant problem since the bulk cosmological constant controls the flatness of the solutions and should be set to zero in order to obtain Minkowski four dimensional spacetime. 

Nevertheless, if one accepts to do only one fine-tuning, the one of the bulk cosmological constant, the above self-tuning mechanism can guarantee flatness of the solutions irrespective of the brane tension. Then the challenge that is posed is to find some dynamics which can relax the bulk cosmological constant to zero in order to avoid the latter fine-tuning.

\vskip1cm

\textbf{Acknowledgments:} We would like to acknowledge helpful discussions with Stefan F\"orste, Hans-Peter Nilles and Gianmassimo Tasinato. This work is supported in part by the European Community's Human Potential
Programme under contracts HPRN--CT--2000--00131 Quantum Spacetime,
HPRN--CT--2000--00148 Physics Across the Present Energy Frontier
and HPRN--CT--2000--00152 Supersymmetry and the Early Universe. H.M.L. was also supported by priority grant 1096 of the Deutsche
Forschungsgemeinschaft.

\vskip1cm

\textbf{Note added:} While this work was being completed, \cite{Rubakov} appeared in the pre-print archives considering the same spherical sigma model and obtaining similar solutions.

\end{document}